\RequirePackage{fix-cm}
\documentclass[smallextended]{svjour3}       
\smartqed  
\usepackage{graphicx}
\usepackage{multirow}
 \journalname{DOI: 10.1007/s11038-020-09533-9 Earth, Moon and Planets}
\begin{document}

\title{Statistical characteristics on SEPs, radio-loud CMEs, low frequency type II and type III radio bursts associated with impulsive and gradual flares}

\titlerunning{Statistical characteristics on SEPs radio-loud CMEs  of type II \& type III}        

\author{ P. Pappa Kalaivani{$^1$$^*$} \and
        A. Shanmugaraju{$^2$} \and
        O. Prakash {$^3$$^,$$^4$$^*$}\and 
        R.-S. Kim{$^5$$^,$$^6$}
}


\institute{P. Pappa Kalaivani \at
              {$^1$}Department of Physics, Ultra College of Engineering \& Technology for Women, Ultra Nagar, Madurai -- 625 104, Tamil Nadu, India\\
              {$^*$}\email{kalai2230@gmail.com}\\          
           \and
            A. Shanmugaraju \at
            {$^2$}Department of Physics, Arul Anandar College, Karumathur, Madurai -- 625514, Tamil Nadu, India\\
            \and
            O. Prakash \at
            {$^3$}Department of Physics,Sethu Institute of Technology, Pulloor, Kariapatti,Viruthunagar, Tamil Nadu -- 626 115, India\\
            {$^4$}Key Laboratory of Dark Matter and Space Astronomy, Purple Mountain Observatory, Chinese Academy of Sciences, Nanjing -- 210008, Jiangsu, China\\
             {$^*$}\email{prakash18941@gmail.com} 
            \and
            R.-S. Kim \at
            {$^5$}Korea Astronomy and Space Science Institute, Daejeon, Korea\\
            {$^6$}University of Science and Technology, Daejeon, Korea
            }
            
\date{Received: 19 Sep 2019 / Accepted: 07 Jul 2020}

\maketitle

\begin{abstract}
We have statistically analyzed a set of 115 low frequency (Deca-Hectometer wavelengths range) type II and type III bursts associated with major Solar Energetic Particle (SEP: E$_p$ $>$ 10 MeV) events and their solar causes such as solar flares and coronal mass ejections (CMEs) observed from 1997 to 2014. We classified them into two sets of events based on the duration of the associated solar flares:75 impulsive flares (duration $<$ 60 min) and 40 gradual flares (duration $>$ 60 min).On an average, the peak flux (integrated flux) of impulsive flares X2.9 (0.32 J m$^{-2}$) is stronger than that of gradual flares M6.8 (0.24 J m$^{-2}$). We found that impulsive flare-associated CMEs are highly decelerated with larger initial acceleration and they achieved their peak speed at lower heights (--27.66 m s$^{-2}$ and 14.23 R$_o$) than the gradual flare-associated CMEs (6.26 m s$^{-2}$ and 15.30 R$_o$), even though both sets of events have similar sky-plane speed (space speed) within LASCO field of view. The impulsive flare-associated SEP events (Rt = 989.23 min: 2.86 days) are short lived and they quickly reach their peak intensity (shorter rise time) when compared with gradual flares associated events (R$_t$ =1275.45 min: 3.34 days). We found a good correlation between the logarithmic peak intensity of all SEPs and properties of CMEs (space speed: cc = 0.52, SE$_cc$ = 0.083), and solar flares (log integrated flux: cc = 0.44, SE$_cc$ = 0.083). This particular result gives no clear cut distinction between flare-related and CME-related SEP events for this set of major SEP events. We derived the peak intensity, integrated intensity, duration and slope of these bursts from the radio dynamic spectra observed by Wind/WAVES. Most of the properties (peak intensity, integrated intensity and starting frequency) of DH type II bursts associated with impulsive and gradual flare events are found to be similar in magnitudes. Interestingly, we found that impulsive flare-associated DH type III bursts are longer, stronger and faster (31.30 min, 6.43 sfu and 22.49 MHz h$^{-1}$) than the gradual flare- associated DH type III bursts (25.08 min, 5.85 sfu and 17.84 MHz h$^{-1}$). In addition, we also found a significant correlation between the properties of SEPs and key parameters of DH type III bursts. This result shows a closer association of peak intensity of the SEPs with the properties of DH type III radio bursts than  with the properties DH type II radio bursts, at least for this set of 115 major SEP events.
\end{abstract}

\section{Introduction}
\label{intro}
The solar energetic particle (SEP) events are often tend to be accelerated by the solar flares and coronal mass ejections-driven shocks \cite{Kahler1978} \cite{Reames2004}\cite{Emslie2012}. These fast moving energetic protons and associated magnetic field are the main cause for geo and space-weather effects such as electronic damages on satellites \cite{MacAlester2014}, radian hazards for polar flights, astronauts, and outages of the power grids \cite{Bothmer2007}.The prolonged acceleration of the SEPs mainly depends on different types of accelerating mechanism, source locations and free available magnetic energy \cite{Engell2017}\cite{Swalwell2018}. It is generally accepted that the short duration SEP events are accelerated by solar flares \cite{Mason1999}\cite{Reames1999}, and the long duration SEP events are related with CMEs-driven shocks \cite{Kocharov2002}. But most of the halo CMEs is found to be associated with stronger solar flares \cite{Harrison1995} \cite{Aarno2011} which produce the higher energetic particle events and magnetic storms. \cite{Kahler1996} found that the peak intensity of SEP events is significantly correlated with CME-driven shocks speed. Scattering of data points in this plot indicated the involvement of other factors in the particle acceleration \cite{Kahler2001} \cite{Gopalswamy2003} \cite{Cliver2006} \cite{Mewaldt2012}. Later many authors found that the peak intensity of SEPs correlated with both speed and width of the associated CMEs \cite{Kahler1984} \cite{Reames1999} \cite{Gopalswamy2003} \cite{Papaioannou2016} \cite{Ding2013} \cite{Prakash2017}. Significant number of authors pointed out that duration of the solar flares is important parameter to predict the space weather events \cite{Cane1986} \cite{Kallenrode1992} \cite{Swalwell2018}. \cite{Cane1986} divided the SEP associated solar flares into two groups: impulsive and gradual (long duration) solar flares during 1978 -- 1983. In the gradual flares, the peak intensities still remain 10\% for more than 60 min, whereas in the impulsive flares peak intensities return to below the threshold level within 60 min. They found that impulsive flares are associated with meter wavelength type III radio bursts, but not associated with interplanetary shocks. But the gradual flares originated anywhere on the solar disk were found to be associated with coronal and interplanetary shocks. Only half of the events were associated with meter wavelength type III radio bursts. Furthermore, many authors suggested that the ambient suprathermal seed particles from the previous solar flares \cite{Mason1999} \cite{Cane2006} \cite{Ding2015} and from the preceding CMEs \cite{Gopalswamy2002a} \cite{Li2012} \cite{Ding2015} \cite{Lawrance2016} \cite{PappaKalaivani2019} can play an important role for producing the large SEP events.\\

\indent When a CME is considerably faster than the local Alfven waves, it is able to produce the magneto hydrodynamic (MHD) shocks propagating outward through the solar atmosphere \cite{Nelson1995}. These shocks can be observed as type II radio bursts from high to low frequencies, because the electron density decreases radially outward from the solar surface. These CME-driven shocks are also believed to accelerate protons to larger energies. The type II radio emissions in metric (m) and deca-hecto metric (DH) wavelength are detected by the ground and onboard radio spectrographs, respectively. The m type II bursts are observed in the frequency range of 300 MHz --18 MHz by various ground based radio solar telescope network (RSTN). Due to the ionospheric cutoff, the radio emission of DH and kilometric type II bursts should be observed in the frequency range approximately from 16 MHz to 20 kHz by the WAVES instruments on the Wind and SWAVES onboard STEREO spacecraft. On the other hand, type III radio emission is a fast drifting feature, usually originated from the electrons accelerated along the open magnetic field lines during the solar flares \cite{Wild1950} \cite{Nicholson1978}.\\

\indent The intense long duration type III radio bursts were analyzed using the data observed from the ISEE-3 spacecraft \cite{Cane1981}. \cite{MacDowall1987} statistically examined the long duration type III bursts associated with shock associated (SA) events. \cite{Cane2002} studied the type III bursts in the coronal and interplanetary (IP) space using the ground based spectrograph and WAVES data. They reported that these type III events are always associated with energetic CMEs and they can be a new source of electron acceleration. Many authors found that the large gradual SEP events are associated with DH type II bursts \cite{Gopalswamy2002b}\cite{Laurenza2009} \cite{Cliver2009}. Recently \cite{Winter2015} extensively analyzed a set of 123 DH type II radio bursts from the dynamic spectra observed by the Wind/WAVES and their associated SEP and non-SEP associated events during 2010--2013. They demonstrated that high degree of peak intensity of SEP events is associated with both DH type II and DH type III radio bursts. To forecast the SEP events, they also computed a new radio index from the principle component analysis using the derived properties of DH type III and DH type II radio bursts. They pointed out that peak intensity and duration of DH type III radio bursts could also be the dominant factors along with the properties of DH type II radio bursts. Finally, they concluded that DH type III bursts observed along with DH type II bursts play an important role in forecasting the SEP events. Hence a comprehensive analysis of the radio--loud SEP events and their solar causes (flares and CMEs) covering a longer period would lead to more understanding of the relation between these events. Also, \cite{Kouloumvakos2015} demonstrated that the properties of SEP events can be inferred from their associated solar type II and type III radio emissions during Solar Cycle 23 for selected 115 high energy proton ($\sim$68 MeV) events. In addition to this, they provide the spatial and temporal information about the particle release time and their heights. They clearly pointed out that most of SEP events are found to be associated with both type II and type III radio bursts, but a good association percentage is obtained only in cases accompanied by type III radio bursts not in the cases of type II bursts. They concluded that both solar flares and CME-associated shocks related SEP events are observed in the major proton events ($>$ 50 MeV). Finally, they indicated that a clear-cut distinction between flare-related and CME-related SEP events is difficult to establish. \\

\indent Main purpose of the present paper is to address the following questions: Whether major SEP events are related to both solar flares and CMEs or not, and what kind of basic characteristics of DH type II and DH type III bursts are related to these SEP events. For this work, we concentrate on the basic characteristics and their relationships between the solar flares, CMEs and SEP events based on the classification of impulsive and gradual flares associated events observed from 1997 to 2014 in Solar Cycle 23 and 24 (SC 23 and 24). We also derive the basic attributes of DH type II and DH type III bursts (associated with solar flares and CMEs, major SEP events) from the dynamic spectra observed by WAVES instrument and compare them based on two sets of solar flares (impulsive and gradual). Furthermore, we have analyzed the role of DH type II and DH type III associated shock strength and speeds for particle acceleration. In Section 2, we describe the method of selection of events and analysis. In Sections 3 and 4, SEP characteristics according to impulsive and gradual flare events are presented. A brief summary and conclusion are presented in Section 5.
\section{Data analysis and Event selection}
\label{sec:2}
\indent We have utilized 143 major SEP events (E$_p$ $>$ 10 MeV) and their associated solar flares, CMEs and DH type II bursts listed in the CDAW\'s (Coordinated Data Analysis Workshops) data Center\footnote{$https://cdaw.gsfc.nasa.gov/CME_list/sepe/$} during the period November 1997 -- December 2014. From these events, we have adopted 115 flare events which are well observed and clearly located. They are classified into two sets of events based on the duration of solar soft X-ray flares (1-- 8 A) observed by GOES (Geostationary Operational Environmental Satellite): i) explosive or impulsive flare events having duration less than 60 min and ii) progressive or gradual flare events having duration longer than 60 min (Long Duration: LD). Flare duration is considered from the onset to the decay of the flux to half of the peak intensity of the corresponding flare. We adopted the flare onset time as in the CDAW\'s major SEP event list. CDAW\'s data team has made the onset correction of solar flares by taking the time derivative of soft X-ray longer wavelength flux (shown in Figure 1a and 1b). We carefully checked the source location and active region (AR) from GOES soft X-ray flare data\footnote{$https://www.ngdc.noaa.gov/stp/space-weather/solar-data/solar-features/solar-flares/x-rays/goes/xrs$} as a reference of each event. We adopted the end time and integrated flux (those are not listed in major SEP events list) from the GOES soft X-ray flare data. In addition to CDAW\'s major SEP event list, we have compiled the other important key parameters for solar flares, CMEs and major SEP events. The residual acceleration of the CMEs are taken from the SOHO/LASCO (Solar and Heliospheric Observatory/Large Angle and Spectrometric Coronagraph) CME catalog\footnote{$https://cdaw.gsfc.nasa.gov/CME_list$} maintained by the NASAs CDAW data Center \cite{Yash2004} \cite{Gopalswamy2009}. For halo CMEs, the space speed is (projection corrected speed) taken from the CDAW halo CME list\footnote{$https://cdaw.gsfc.nasa.gov/CME_list/halo/halo.html$}, which was derived from the sky-of-plane speed and longitude of source location by using the Cone model \cite{Xie2004}. The geometrical correction was made for non-halo CMEs by multiplying the derived sky-plane speed by the factor of 1/sin $\theta$, where $\theta$ is the angular distance of the CME source from the solar disk center. Furthermore, we derived the initial and final speeds of SEPs-associated CMEs from the first and last two SOHO/LASCO height-time (h-t) measurements, respectively. We also estimated the peak speed and their corresponding peak speed height of the CMEs from the height time measurement. To obtain the properties of major SEP events, we have downloaded the 5-minute average integral proton flux from the GOES west facing instruments (GOES-8, 11, 15)\footnote{$https://sohoftp.nascom.nasa.gov/sdb/goes/particle/$}. For each event, we plotted the flux profile and verified the start time, peak time, and peak flux of SEP events listed in the CDAW?s major SEP event list. In addition to this, end time of major SEP events is determined when proton flux ($>$ 10 MeV) falls below 1 pfu (1 pfu =1protons cm$^{-2}$ sr$^{-1}$ s$^{-1}$) on the other side of the peak. Using the start and end times, we found the duration and integrated intensity for all major SEP events. But for few events, ending times were not determined because before the end of the previous SEP event, another particle event had started.\\
\indent In order to illustrate impulsive and gradual flare events, the soft X-ray solar flare light curves at 1-- 8 A (red) and 0.5 -- 4 A (blue) from GOES observations are plotted in Figure 1 (top). The bottom panels show the time derivative of the flux in 1 -- 8 A for impulsive and gradual flare events. The M5.1 class impulsive flare shown in Figure 1a was observed on 17-May-2012 at N11W76 with start, peak and end times as 01:25 UT, 01:47 UT, and 02:14 UT, respectively. These details are also seen from the time derivative of the soft X-ray curve shown in Figure1a (bottom). The duration of this solar flare was nearly 49 min and hence considered as an impulsive event. This impulsive flare-associated halo CME was first observed by the LASCO/C2 coronagraph at a height of 3.61 Ro around 01:48 UT. The mean sky-plane speed and acceleration was reported as 1582 km s$^{-1}$ and -51.8 m s$^{-2}$, respectively. On 31-August-2012, a long duration C8.1class flare originated from N25E59 at 19:45 UT. The corresponding flare peak and end times were reported as 20:43 UT and 21:51 UT, respectively. The duration of this solar flare was around 133 minutes and hence considered as gradual flare event. This duration of this event was also checked from the time derivative of flare flux in 1-- 8 A and plotted in Figure1b (bottom). This gradual flare- associated halo CME erupted at 20:00 UT during the rising phase of the gradual flare and first observed by the LASCO/C2 at a height of 3.01 R$_o$. The means sky-plane speed and residual acceleration of associated CME were 1442 km s$^{-1}$ and 2 m s$^{-2}$, respectively within the LASCO FOV.\\

\indent In addition to the characteristics of parent solar eruptions and their associated SEP events, our main concern is to derive the key aspects of DH type II and DH type III bursts. We have carefully compiled the end time, starting frequency and ending frequency of DH type II radio bursts from the online CDAW\'s Wind/WAVES catalogue. Please note that the association between the solar flares, major SEPs, CMEs, DH type II radio bursts was followed as given in the CDAW\'s major SEP event list. The other important basic properties (peak intensity, integrated intensity, duration and slope which are not listed in the online Wind/WAVES catalogue) of DH type II and DH type III radio bursts are derived from the actual radio dynamic spectra. Recently, \cite{Winter2015} derived the same properties for a different set of DH type II and DH type III radio bursts during 2010 -- 2013. We also used the same method to analyze the sample of 115 Wind/WAVES dynamic spectra for determining the properties of DH type II and DH type III bursts. The DH type II and DH type III bursts are recorded by the three detectors RAD1 (20 -- 1040 kHz), RAD2 (1.075 -- 13.825 MHz) and TNR (4 -- 245 kHz) which constitue the Radio and Plasma Wave (WAVES) experiment \cite{Bougeret1995} in the Wind spacecraft \cite{Acuna1995}. From these three detectors, the calibrated one-minute averages IDL save files were downloaded for each sample of event from the Wind/WAVES. In each IDL save file, the ratio (R) and  the background values (B) are recorded and stored in the units of µV Hz $^{-1/2}$ \cite{Hillan2010}.  These R and B values are converted into solar flux units (1 sfu = 10$^{-22}$ W m$^{-2}$ Hz$^{-1}$) using the relation J (sfu) = 10$^{10}$ (R$\times $B)/(Z$_o$ $\times$ A), where A is the area of the antenna in the unit of m$^2$ (RAD1 has an area of 1225 m$^2$) and Z$_o$ is the impedance of free space (377 Ohms). Before plotting, we made some corrections like removing the gaps from the detector; boxcar smoothing and interpolating the R value to remove zeros, particularly in RAD1. We have plotted the dynamic spectra for the whole time periods for each of the 115 radio bursts, plotted and picked up local peak intensity at each frequency. Each data point was fitted along with the linear function and then these data are transformed into 1/frequency-space for finding the slope of DH type II radio bursts.  We also estimated the integrated intensity along the fitted line between the points where the flux falls to 15\% of the local peak intensity.  In the same way, we also determined the peak intensity, integrated intensity, slope of DH type III radio bursts and the duration of the DH type III burst using the time where 1 MHz signal exceeds 6 dB or four times of background \cite{MacDowall2003}.

\begin{figure*}[!ht]
\setlength \abovecaptionskip {0.5 \baselineskip}
\setlength \belowcaptionskip {0pt}
\begin{center}
  \includegraphics[width=0.95 \textwidth]{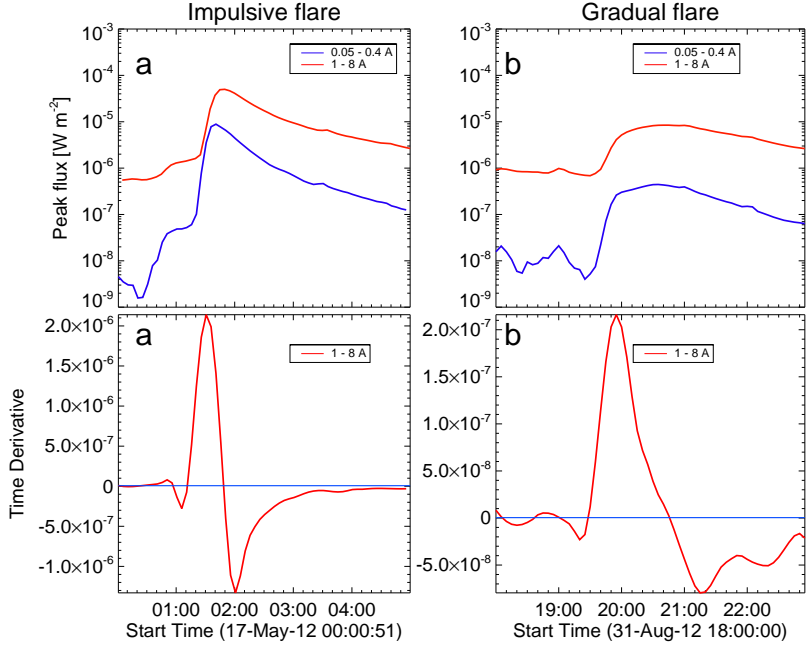}
\caption{Examples of (a) impulsive, and (b) gradual flare events. The top panels present the soft X-ray solar flare flux observed by GOES at 1-8 A (red) and 0.5- 4 A (blue). The bottom panels show the time derivative for the flux 1- 8 A in which the horizontal solid blue line represents the zero of the flux time derivative for impulsive and gradual flare events.}
\label{Figure 1}
\end{center}       
\end{figure*}

The mean, median, standard statistical error (SE) and probabilistic Student t-test values of the key parameters of solar flares, CMEs, major SEP events, DH type II bursts and DH type III radio bursts are presented in Table 1 separately for impulsive and gradual flares associated events. In Column 1 the name of event is listed.  Column 2 denotes the basic attributes of the corresponding events. Columns 3 and 4 give the estimated mean (median) and standard error (SE) of the considered parameters for impulsive flare associated events and Columns 5 and 6 present the mean (median) and standard error (SE) for gradual flares associated events. The standard statistical error has been estimated from the standard deviation (SD) and number of events $(SD/\sqrt(n))$ where n is the number of data point. Column 7 presents the probability Student t-test value. We determined the Student t-test value by assuming that the two populations have normal distributions with the assumptions of an equal variance.

\begin{table}[htp]
\caption{Mean, median, standard statistical error (SE) and probability test values of properties of solar flares, CMEs, SEPs, DH type II, and DH type III radio bursts for impulsive and gradual flare associated events}
\begin{center}
\begin{tabular}{|c|l|c|c|c|c|c|}
\hline

\multirow{2}{*}{Events} &\multirow{2}{*}{Properties}& \multicolumn{2}{c}{Impulsive events}& \multicolumn{2}{c} {Gradual events} & \multirow{2}{*}{t-test P value}\\
& &Mean (Median)&SE& Mean (Median)& SE&  \\\hline 

\multirow{2}{*}{Flare} & Peak Flux (W m$^{-2}$)& X2.9 (M9.8)& B6.5& M6.8 (M4.15)& B2.0& 5\%\\
&Integrated flux (J m$^{-2}$)& 0.32 (0.12)&0.06& 0.24 (0.15)& 0.04& 36\%\\\hline
\multirow{9}{*}{CME} &Sky-plane speed (km s$^{-1}$)& 1575.94 (1481.50)& 71.09&1571.72 (1494.00)& 78.52& 98\%\\

&Space speed (km s$^{-1}$)&1779.41 (1618.58) &79.30&1771.94 (1631.00)&87.59&95\%\\
&Acceleration (m s$^{-2}$) &-27.66 (-21.65) &7.84 &6.26 (2.00) &5.66 &2\%\\

&Initial acceleration (km s$^{-2}$)
&2.01 (1.48)
&0.16
&0.67 (0.65)
&0.06
&0.001\%\\

&Initial speed (km s$^{-1}$)
&1738.09 (1672.33)
&81.70
&1429.34 (1326.75)
&94.98
&1\%\\

&Final speed (km s$^{-1}$)
&1601.02 (1485.44)
&93.03
&1647.40 (1504.69)
&92.67
&74\%\\

&Peak speed ( km s$^{-1}$)
&1972.68 (1856.00)
&89.93
&1940.22 (1756.00)
&98.99
&81\%\\

&Peak speed height (R$_o$)
&14.23 (12.13)
&0.97
&15.30 (17.30)
&1.19
&49\%\\

&Width (deg)
&336.18 (360)
&9.22
&329.36 (360)
&8.59
&61\%\\\hline

\multirow{4}{*}{SEP}
&Rise time (min)
&989.23 (755.00)
&99.35
&1275.45 (1012.50)
&168.31
&12\%\\

&Duration (days)
&2.86 (2.60)
&0.22
&3.34 (2.91)
&0.34
&21\%\\

&Log peak intensity (pfu)
&2.22 (1.97)
&0.10
&2.11 (1.90)
&0.14
&52\%\\

&Log integrated flux (pfu-min)
&4.38 (4.21)
&0.12
&4.24 (4.23)
&0.14
&44\%\\\hline
\multirow{5}{*}{DH type II}
&Duration (min)
&856.87 (450)
&110.60
&1027.23 (870)
&140.82
&35\%\\

&Ending frequency (kHz)
&1197.61 (200)
&263.53
&550.21 (150)
&169.73
&8\%\\

&Log peak intensity (sfu)
&3.64 (3.52)
&0.09
&3.65 (3.67)
&0.10
&97\%\\

&Log integrated intensity (sfu-min)
&7.34 (7.32)
&0.07
&7.30 (7.24)
&0.09
&74\%\\

&Slope (MHz h$^{-1}$)
&0.72 (0.64)
&0.04
&0.71 (0.61)
&0.06
&94\%\\\hline

\multirow{4}{*}{DH type III}
&Duration (min)
&31.29 (31.00)
&1.21
&25.08 (25)
&2.17
&0.1\%\\

&Log peak intensity (pfu)
&6.43 (6.51)
&0.08
&5.85 (5.86)
&0.10
&0.0002\%\\

&Log integrated intensity (pfu-min)
&10.05 (10.12)
&0.67
&9.45 (9.42)
&0.74
&0.0001\%\\

&Slope (MHz h$^{-1}$)
&22.49 (19.47)
&1.14
&17.84 (15.35)
&1.23
&1\%\\\hline

\end{tabular}
\end{center}
\label{default}
\end{table}%

\section{Characteristics of radio-loud CMEs, flares and SEPs of impulsive and gradual events}
\subsection{Characteristics of impulsive and gradual flares}
Figure 2 presents the distributions of peak flux and integrated flux (fluence) of all major SEP associated flares (top), impulsive flares (middle) and gradual flares (bottom). In general, many (53\%) of the major SEP associated events have M class flares and the remaining 37\% and 10\% of events are associated with X and C class flares, respectively. It should be noted that 13\% and 55\% of major SEP associated events of impulsive and gradual flares belong to X class events, respectively. Student t-test shows the difference in the mean peak flux of impulsive and gradual flares (X2.9 and M6.8, respectively) as statistically significant (P = 5\%). The estimated standard statistical error for impulsive flare associated events (SE = B6.5) is slightly larger than that of gradual flare associated events (B2.0). Furthermore, integrated flux for impulsive flare events (0.32 J m$^{-2}$) is larger than that of gradual flares (0.24 J m$^{-2}$), and these means have statistically insignificant (P = 36\%) difference. Here, we also have found that the same tendency of larger estimated SE for impulsive flare events (0.06 J m$^{-2}$) than the gradual flare events (0.04 J m$^{-2}$).

\begin{figure*}[!ht]
\setlength \abovecaptionskip {0.5 \baselineskip}
\setlength \belowcaptionskip {0pt}
\begin{center}
  \includegraphics[width=0.95 \textwidth]{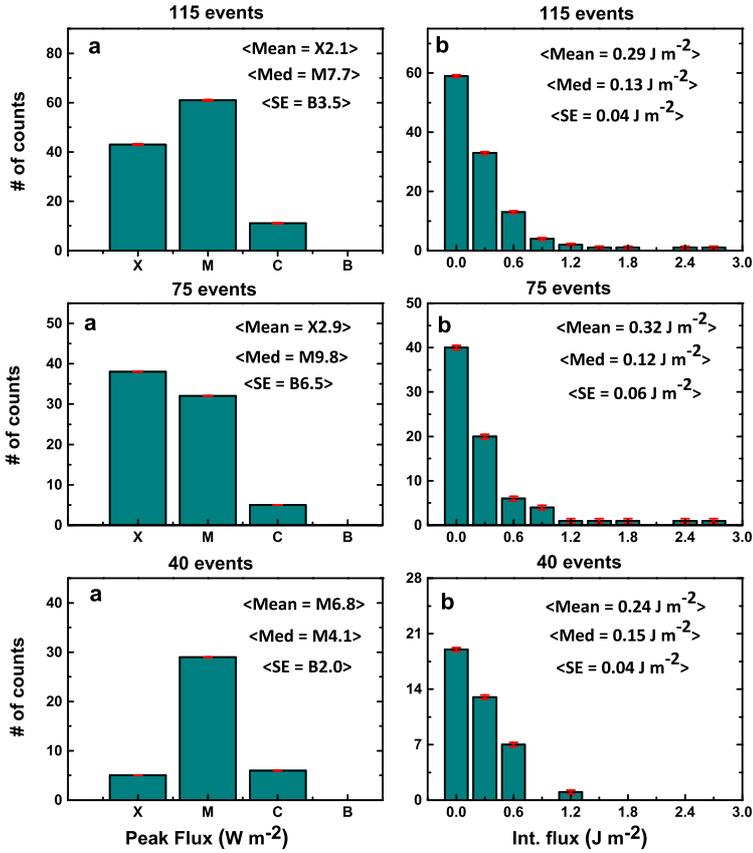}
\caption{Distributions of: (a) peak flux; and (b) integrated flux (fluence) for all major SEP- associated flares (top), impulsive (middle) and gradual flare (bottom) events. The error bars denote the standard statistical error (SE).}
\label{Figure 2}       
\end{center}
\end{figure*}

\subsection{Source Distribution}
Generally, most of shock-associated SEP events are observed at a wide range of longitudes \cite{Rouillard2012} \cite{Lario2016}. As seen from Figure 3, we observed the same distribution for the present set of 115 major SEP events during solar cycles 23 and 24. Many works emphasized that heliolongitude of parent solar eruptions (solar flares and/or CMEs) are an important factor to characterize the associated major SEP events \cite{Van1975} \cite{Cane1988} \cite{Kahler2007} \cite{Balch2008}. Recently, \cite{Park2010} explained the significance of solar longitude for a set of 166 proton events associated with major flares during 1976 -- 2006. They indicated that the high degree of the occurrence rate of SEP events was from the westward events. But, they found that peak fluxes of eastern side originated solar flares are well correlated with peak intensity of associated SEP events. However, for the convenience, we also classified our sample of events into three categories based on the longitudes (L) irrespective of latitudes: i) eastern side (L $>$ E30$^{o}$); ii) disk center (E30$^{o}$ $\leq$  L $\leq$ W30$^{o}$); and iii) western side (L $> $W30$^{o}$). The triangle symbols in Figure 3 represent the location of the impulsive flares associated SEP events and asterisk symbols represent the gradual flares associated events. 

\begin{table}[htp]
\caption{Solar proton events occurrence probability depending on the three different longitudes for impulsive and gradual events}
\begin{center}
\begin{tabular}{|c|c|c|c|c|c|c|}
\hline

\multirow{2}{*}{Properties}&
\multicolumn{3}{c}{Impulsive flare events (75)} &
\multicolumn{3}{c}{Gradual flare events (40)}\\

&East (9)
&Center (25)
&West (41)
&East (7)
&Center(12)
&West (21)\\\hline

Space Speed (km s$^{-1}$)
&1738.57
&2274.83
&1637.35
&1760.70
&1881.17
&1727.23\\

SEP Peak Intensity (pfu)
&472.08
&4567.65
&335.64
&72.49
&1198.43
&1394.61\\\hline

\end{tabular}
\end{center}
\label{default}
\end{table}%

Out of 115 events, only 14\% and 32\% of major SEPs associated source locations are distributed in the eastern and disk center respectively and the remaining 54\% of events are from the western side.  The mean space speed of all 115 CMEs originated from the eastern side (1749.64 km s$^{-1}$) is almost equal to events from the western side (1682.29 km s$^{-1}$). But the mean peak intensity of SEPs from western side (865.13 pfu) is three times larger than events from eastern side (272.29 pfu). Furthermore, mean space speed of CMEs and peak intensity of SEPs originated from disk events are 2078 km s$^{-1}$ and 2883.04 pfu, respectively. In the aspect of impulsive flare and gradual flare associated events, 12\% and 18\% of impulsive and gradual flares associated SEP events were observed in the eastern side as given in Table 2. The average peak intensity of impulsive flare events (472.08 pfu) is larger than that of gradual flares (72.49 pfu) associated SEP events, even though both sets of events have similar mean space speeds. But this scenario in SEP peak intensity completely reverses in the western side events. i.e., Even though both sets of events (impulsive and gradual) have similar mean space speed, and almost equal number (55\% and 53\%) of events originated from the western side, the peak intensity of gradual flare associated SEP events (1394.61 pfu) is significantly larger than that of impulsive flares associated events (335.64 pfu). \cite{Kouloumvakos2015} pointed out that almost 69\% of the flares associated SEP events were originated from the western hemisphere. Our result is also consistent with that of \cite{Kouloumvakos2015}.  From the Student t-test, the difference in the mean of the peak intensity of SEPs is found to be statistically significant (P = 4\%). \\
\indent Only 33\% and 30\% of impulsive flare and gradual flare associated events, respectively, are detected from the disk center. On an average, space speed of CMEs for impulsive flare associated event (2274.83 km s$^{-1}$) is larger than the gradual flare associated events (1881.17 km s$^{-1}$). Further, impulsive flares associated SEP peak intensity (4567.65 pfu) is three times larger than that of gradual flares associated SEP events (1198.43 pfu). From these results we infer that, magnetically well connected (western) gradual flare associated SEP events are more energetic than the impulsive flares associated events.\\

\begin{figure*}[!ht]
\setlength \abovecaptionskip {0.5 \baselineskip}
\setlength \belowcaptionskip {0pt}

  \includegraphics[width=0.95 \textwidth]{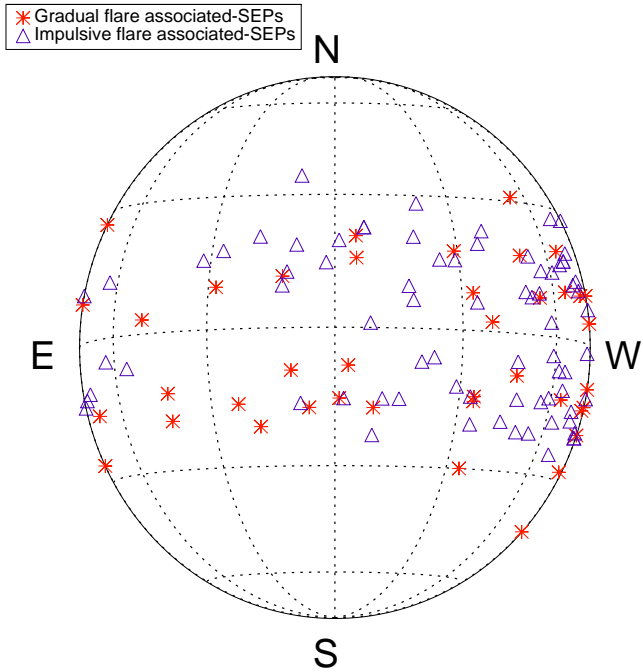}
\caption{Source distributions of all flares associated with major SEP events. Triangle symbols represent the locations of impulsive flare associated events. Asterisk symbols give the locations of gradual flare associated events.}
\label{Figure 3}       
\end{figure*}

\subsection{Characteristics of radio-loud CMEs}
We briefly discuss the various key parameters of major SEP associated CMEs, mostly adopted from SOHO/LASCO CME catalog. The distributions of initial speed (a), final speed (b), residual acceleration (c), and initial acceleration (d) for all major SEP associated CMEs (top row), impulsive flare associated CMEs (middle row) and gradual flare associated CMEs (bottom row) are plotted in Figure 4. The initial and final speeds derived from the first and last two h?t measurements observed by the LASCO FOV are shown in Figure 4a and 4b, respectively. Initial speed of almost 92\% of major SEP associated CMEs is larger than 1000 km s$^{-1}$. On an average, the initial speeds of impulsive flare associated CMEs (1738.09 km s$^{-1}$; SE = 81.70 km s$^{-1}$) are considerably larger than that of gradual flare associated CMEs (1429.34 km s$^{-1}$; SE = 94.98 km s$^{-1}$). The Student t-test shows that the difference in these two values is statistically significant (P = 0.01\%). On the other hand, we found that the average and median final speeds of the both sets of flare associated CMEs are almost similar as shown in Figure 4b. From this result we infer that impulsive flare associated CMEs are highly accelerated (associated with stronger flares) in the lower corona than that of gradual flare associated CMEs. \\
\indent Figure 4c shows the distributions of mean acceleration or residual acceleration for major SEP associated CMEs (top), impulsive flares associated CMEs (top) and gradual flare associated CMEs (bottom). The acceleration due to gravity and the propelling forces decrease significantly at heights corresponding to the LASCO FOV, where the deceleration due to drag becomes dominant (so-called residual acceleration). We have only considered the well observed CMEs with more than three h-t measurements. On an average, major SEP associated CMEs are considerably decelerated -- 14.99 m s$^{-2}$ and it is nearly three times greater than the average residual acceleration of 438 limb CMEs reported by \cite{Gopalswamy2012a}. Interestingly, it is found that impulsive flares associated CMEs are more decelerated (-- 27.66 m s$^{-2}$) than the gradual flares associated CMEs (6.26 m s$^{-2}$) within the LASCO FOV and this mean difference is statistically significant (P = 2\%). The residual acceleration of the impulsive flare associated CMEs (SE = 7.84 m s$^{-2}$) are found to have a relatively larger statistical error than that of gradual flare associated events (SE = 5.66 m s$^{-2}$). This is consistent with the results of \cite{Jang2017} that the gradual flare associated CMEs have longer accelerating phase due to the longer duration of magnetic reconnection than that of impulsive flare associated CMEs.

\begin{figure*}[!ht]
  \includegraphics[width=0.95 \textwidth]{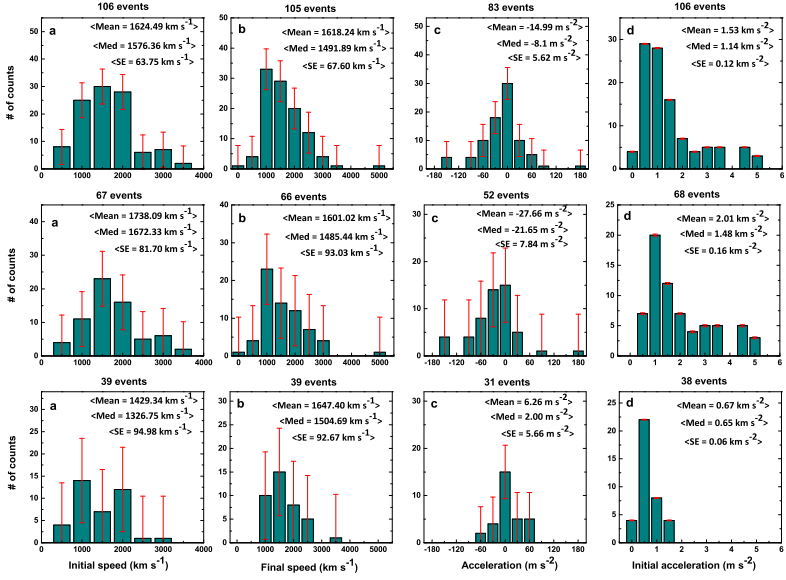}
\caption{Distributions of: (a) initial speed, (b) final speed, (c)  acceleration, and (d) initial acceleration for all major SEP  associated CMEs (top), impulsive flares associated CMEs (middle) and gradual flares associated CME (bottom) events. The error bars denote the standard statistical error (SE).}
\label{Figure 4}       
\end{figure*}

\indent The estimation of initial acceleration (ai) of CMEs is a crucial step using the properties of flares and CMEs. As we already discussed, the acceleration of CMEs is mainly due to drag because the propelling force and gravity become negligible in the outer corona and interplanetary space.  But, when CMEs takeoff from the rest in the inner corona (K-corona), initial acceleration has to be positive. We assume that the onset time of CMEs is coinciding with onset of flares and the initial acceleration phase of the CMEs exactly lies during the rising phase of the solar flares \cite{Berkebile2012}. Hence, we have estimated the initial acceleration (a${_i}$ = V${_s}$/T${_f}$) of major SEP associated CMEs using the flare (rise time: T${_f}$) and CME (space speed:V${_s}$) properties. The flare rise time is obtained as the difference between flare peak and flare start times (T${_f}$ = T${_p}$ -- T${_s}$). The distributions of the initial acceleration of major SEP associated CMEs (top), impulsive flares associated CMEs (middle) and gradual flares associated CMEs (bottom) are shown in Figure 4d. The initial acceleration of major SEP associated CMEs has a wide range from 0.08 km s$^{-2}$ to 5.21 km s$^{-2}$. The average initial acceleration (1.53 km s$^{-2}$) of the major SEP associated events is relatively larger than the set of limb DH-CMEs (1.29 km s$^{-2}$) presented in \cite{Prakash2012a}. The average initial acceleration of major SEP impulsive flare associated CMEs (2.01 km s$^{-2}$; SE = 0.16 km s$^{-2}$) is nearly three times larger than the gradual flares associated CMEs (0.67 km s$^{-2}$; SE = 0.06 km s$^{-2}$). The difference has statistically high significance (P = 0.001\%). From the CME observation, many authors have demonstrated that acceleration is a nonlinear function \cite{Wood1999} \cite{Zhang2001}. It is notable that 43\% of impulsive flare associated CMEs have the initial acceleration larger than 2 km s$^{-2}$, but no events have it more than 2 km s$^{-2}$ in the sample of gradual flare associated CMEs. As we discussed earlier, this result is also clearly evident that the larger initial acceleration of impulsive flares associated CMEs should be contributed for the larger initial speeds of the CMEs in the lower corona.\\
\indent The sky-plane speed is the average projected speed of CMEs within the LASCO FOV, derived from the linear fit to the h-t measurements. The mean sky-plane speed ($\sim$1575 km s$^{-1}$) of impulsive and gradual flares associated CME is almost similar for both sets of events (see in Table 1) and the difference in the means is statistically insignificant. However, it is nearly three times larger than that of the average of all CMEs \cite{Gopalswamy2005}. From the LASCO observations, the projected speeds and widths of CMEs on the plane of the sky were derived and catalogued. Hence, real speeds and sizes of CMEs remain unknown due to the projection effect, especially for the disk halo CMEs (longitude $\leq$ 60$_{o}$). \cite{Gopalswamy2000} demonstrated that i) the projection effect is significant for the estimation of the space speed of the halo CMEs and ii) the magnitude of the error depends on the source location of the parent eruptions \cite{Gopalswamy2007}, \cite{PappaKalaivani2010}. In general, the CMEs associated with long wavelength type II radio bursts are relatively energetic than the average CMEs. The space speed of the major SEP associated CMEs is found to vary over a wide range from 633.60 km s-1 to 3682 km s$^{-1}$ with a mean of 1776.66 km s$^{-1}$. Most (36\%) of major SEP associated CMEs have the speeds larger than 1500 km s$^{-1}$. The space speed of CMEs associated with impulsive and gradual flares associated events are almost similar in magnitude, and the difference in the means is statistically insignificant.\\
\indent We also derived the peak speed and their corresponding peak speed height of major SEP associated CMEs (please note that this distribution plot is not presented). The peak height is defined as a height at which the CMEs attained their peak speed within the LASCO FOV. On an average, the impulsive and gradual flares associated CMEs have the similar peak speeds and hence statistical t-test rejects the null hypothesis. However, the impulsive flares associated CMEs (12.13 R$_o$; SE = 0.97 R$_o$) attained their peak speed at lower height than the gradual flares associated CMEs (17.30 R$_o$; SE = 1.19 R$_o$). \cite{Gopalswamy2012b} also studied the peak speed height of the decelerating CMEs and found that these events attained their peak speed at lower heights than the accelerating CMEs. Our result is also consistent with the results of \cite{Gopalswamy2012a} and \cite{Prakash2012a}. Recently, \cite{Jang2017} showed two distinct types of CME-flare relationships based on the observed time differences of flare peak and CME\'s first appearance. They found that if the flare duration is shorter, associated CMEs achieved their peak-speed at lower height.\\
\indent The DH type II bursts are also closely associated with fast and wide CMEs than the average CMEs \cite{Gopalswamy2001b}.Therefore, we examined the widths of the both sets of major SEP associated CMEs and found that the average apparent widths of the impulsive and gradual flares associated CMEs are similar (336.18$^{°}$ and 329.36$^{°}$, respectively; please note that these distributions are also not shown). Almost 83\% and 85\% of impulsive and gradual flares associated CMEs are full halo, and the remaining are partial halo of width from 105$^{°}$ to 315$^{°}$. From the above CME kinematics study, we infer that the impulsive flare associated CMEs lift off with higher initial speeds (propelling force), decelerated more due to drag force and attained their peak speed at lower heights than the gradual flare associated CMEs.

\subsection{SEP characteristics according to the impulsive and gradual flare events}
The distributions of rise time (a), duration (b) and log peak intensity (c) for all major SEP events (top row), impulsive flare associated SEPs (middle row) and gradual flare associated SEPs (bottom row) are presented in Figure 5. The rise time of SEP event is obtained from the time difference between the peak time and onset time, and it gives the degree of impulsiveness of the SEP event. The rise times of 115 major SEP events are in wide range from an hour to 90 hours. The mean (median) rise time of the all major SEP events is 1088.78 min (880 min). The derived standard statistical error is considerably larger for gradual flare associated SEPs (168.31 min) than the impulsive flare associated SEP (99.35 min) events. The Student t-test shows that the difference in the mean rise time of impulsive and gradual flares associated SEP events (989.23 min and 1275.45 min, respectively) is statistically quite significant (P = 12\%).  Almost 57\%and 40\% of impulsive and gradual flare associated SEP events time lower than 1000 min respectively.

\begin{figure*}[h]
  \includegraphics[width=1.0\textwidth]{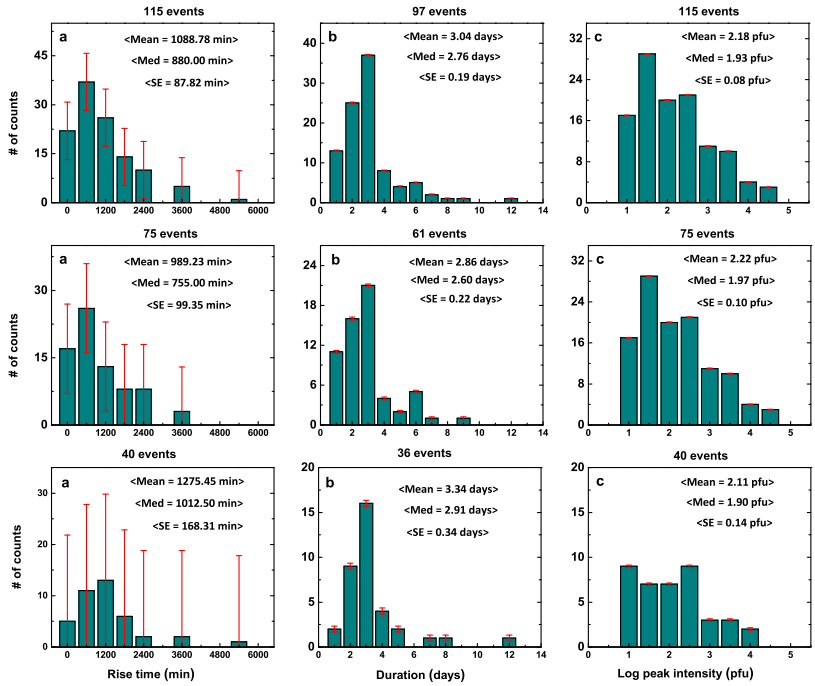}
\caption{Distributions of: (a) rise time; (b) duration; and (c) log peak intensity for all major SEP events (top),impulsive flares associated SEPs (middle) and gradual flares associated SEP (bottom) events. The error bars denote the standard statistical error (SE).}
\label{Figure 5}       
\end{figure*}

\indent In Figure 5b, the distributions of the duration are presented. There is a difference in the mean duration of impulsive and gradual flares associated SEP events (2.86 days and 3.34 days, respectively). Though the difference is statistically insignificant, we infer that the impulsive flare associated SEP events are also more impulsive and short-lived when compared to gradual flare associated SEP events. From the distributions of log peak intensity in Figure 5c, it ranges from 1.04 pfu to 4.50 pfu for major SEP events.  The mean (median) log peak intensity of all major SEP events is 2.18 pfu (1.93 pfu). Most (60\%) of the major SEP events have log peak intensity larger than 2 pfu. However, the difference in the mean log peak intensity of the impulsive and gradual flares associated SEP events (2.22 pfu and 2.11 pfu, respectively) is statistically insignificant (P=52\%). The estimated standard errors ($\sim$0.1 pfu) are also nearly the same. In addition, we also derived the integrated flux for the all major SEP events by integrating the flux along the start and end times of these events. It varies from 2.55 pfu-min to 6.64 pfu-min and the difference in the mean log integrated flux for impulsive and gradual flares associated SEP events (4.38 pfu-min and 4.24 pfu-min, respectively) is statistically insignificant (P = 52\%). These results suggest that the impulsive flare associated SEP events are short-lived, and their peak intensity is achieved earlier than that of gradual flare associated SEPs. 

\subsection{Relationships among the properties of major SEPs, CMEs and solar flares}

Figure 6 presents the relationship between the key parameters of major SEP events and their associated solar flares and CMEs. The correlations between the log peak intensity and CME\'s properties (sky-plane speed and space speed) are shown in Figure 6a-b. The speed of CMEs should be related reasonably well with the peak intensity of the associated SEP events  \cite{Kahler1984} \cite{Kahler2001} \cite{Gopalswamy2003}, as one would expect when the particles are accelerated by the CME-driven shocks. As seen from this Figure 6a (top), we found a positive correlation between log peak intensity and sky-plane speed of CMEs. The estimated Pearson\'s correlation coefficient is cc = 0.45 with the standard error SEcc = 0.086. We estimated the SEcc for each Pearson?s correlation coefficients (SE$_{cc}$) using the relation SEcc =$\sqrt{(1-cc^{2})/(n-2)}$, where n is the number of data points \cite{Papaioannou2016} \cite{Prakash2017}. Generally, when the peak intensity of SEPs is correlated with the CME speed, a high degree of scattering indicates the importance of other factors which contributed for the proton acceleration. Recently, \cite{Papaioannou2016} and \cite{Prakash2017} found that there is a good correlation between log peak intensity and CME speed (cc = 0.57 and cc = 0.58, respectively) for different sets of events and their standard errors are 0.07 and 0.15, respectively (number of data point is also one of the factor to determine the standard error). The correlation between the log peak intensity and sky-plane speed of the CMEs is slightly better for impulsive flare associated events (cc = 0.47; SEcc = 0.107) than the gradual flares associated events (cc = 0.43; SE$_{cc}$ = 0.145). These results are in agreement with the recent studies \cite{Trottet2015} \cite{Dierckxsens2015} \cite{Kim2015}.We also found that there is a better correlation between the log peak intensity and space speed of CMEs (cc = 0.52; SE$_{cc}$= 0.083). After splitting these events as impulsive and gradual flare associated events, magnitude of the Pearson?s correlation coefficient further increases for impulsive flare associated events (cc = 0.55; SE$_{cc}$ = 0.102) but decreases for gradual flare associated events (cc = 0.45; SE$_{cc}$ = 0.144).

\begin{figure*}
  \includegraphics[width=1.0\textwidth]{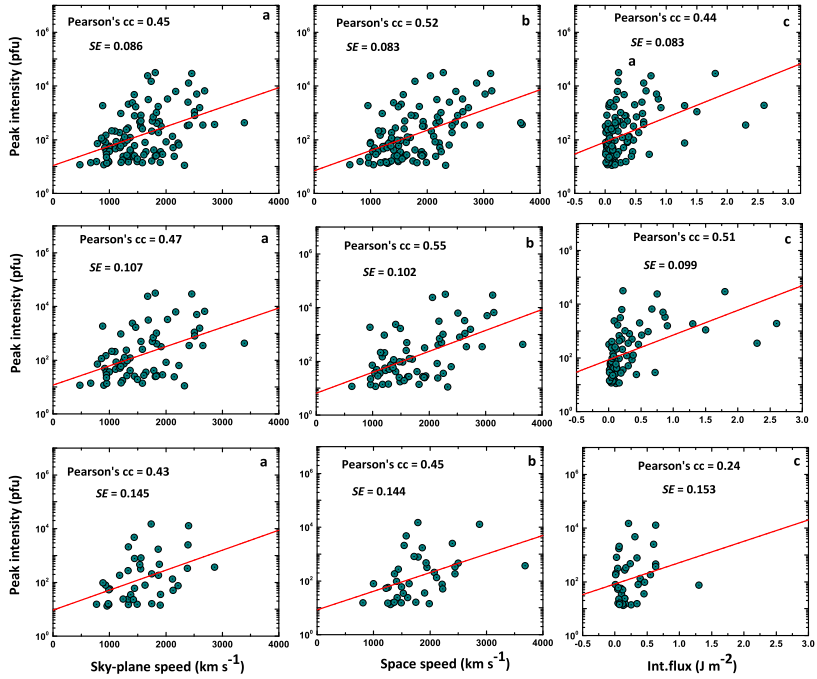}
\caption{Relationship between log peak intensity of SEPs and (a) sky-plane speed, b) space speed of CMEs, and(c) integrated flux of solar flares for all major SEPs (top), impulsive flares associated events (middle) and gradual flares associated events (bottom). Linear fitted functions for data points are shown in red line and the corresponding Pearson?s correlation coefficients cc and SEcc are included in each panel.}
\label{Figure 6}       
\end{figure*}

\indent These results imply that SEP associated CMEs-driven shock is a good electron and proton accelerators. Our results are also consistent with previous studies (e.g., \cite{Gopalswamy2005} \cite{Ding2013} \cite{Prasanna2016}. They suggested that the speeds of CMEs associated with DH type II bursts may influence the peak intensity for a set of SEP events. As mentioned earlier, we checked what other factors contribute to the acceleration of the protons in addition to the speeds of CMEs. The correlation between the log peak intensity and integrated flux of solar flares is plotted in Figure 6c. Interestingly, there is no clear correlation between the log peak intensity of SEPs and integrated flux of flares for gradual flare associated SEPs (cc = 0.24). But, we found a positive correlation for impulsive flare associated SEP events for these properties (cc = 0.51, SEcc = 0.099). Our results are also in support of \cite{Kocharov2017} that the major SEP events previously accelerated by both solar flares and CMEs, later on the particles are accelerated by the CME-driven shocks later. However, our results are also consistent with the results of \cite{Kouloumvakos2015}.  As we discussed earlier, \cite{Kouloumvakos2015} also reported that it is difficult to establish clear cut distinction between flare-related and CME-related SEP events.

\section{DH type II and DH type III radio bursts}
Main concern of the present study is to comprehensively compare the derived properties of DH type II and DH type III bursts for impulsive and gradual flares associated events. Recently, many authors pointed out that DH type II and DH type III radio bursts are potential diagnostics of large SEP events. \cite{Gopalswamy2002a} demonstrated that the brighter and stronger DH type II bursts lanes are observed during the CME-CME interaction events than single CME associated DH type II bursts. Prakash et al. (2017) also pointed out that the CME-CME interaction is important factor for acceleration of large SEP events. Recently, \cite{Winter2015} studied the properties of DH type II and DH type III radio bursts and their correlations with SEP events observed during the 2010 -- 2013. They noted that all 123 SEP events (major SEP events $>$ 10 MeV) were associated with DH type II bursts and 92\% of events were also associated with DH type III bursts. Finally, they concluded that DH type III burst that occur along with a DH type II burst is shown to be an important factor and it can be used to forecast SEP events. In this subsection, we extensively discuss the differences in basic key parameters of DH type II and DH type III radio bursts for impulsive and gradual flares associated events. In general, properties of DH type II radio bursts depend on the kinetic energy of the associated fast and wide CMEs \cite{Gopalswamy2001a} \cite{Gopalswamy2002b} \cite{Gopalswamy2005}.

\subsection{Characteristics of DH type II bursts}

The starting frequency of DH type II bursts may not be considered for indication of SEP events because the ionospheric cutoff frequencies lie on the upper cut-off frequency by Wind/WAVES instruments (14 MHz for Wind). The difference in the mean starting frequencies of DH type II bursts for impulsive and gradual flares associated events (11845.07 and 10835.89 kHz, respectively) is statistically insignificant (P = 25\%). As we discussed earlier, duration and ending frequencies of DH type II bursts are closely related with kinetic energy of associated CMEs. Hence, we studied the ending frequencies of DH type II bursts for both sets of events. We found that 68\% and 71\% of DH type II bursts associated with impulsive and gradual flares were observed below 300 kHz (km domain) and the remaining 32\% and 29\% of events have the emission above 300 kHz. The difference in mean ending frequencies for impulsive and gradual flares associated events (1197.61 and 550.21 kHz, respectively) is quite statistically significant (P = 8\%). The standard statistical error for ending frequency of DH type II bursts for impulsive associated event (SE = 263.53 kHz) is slightly larger than gradual flares associated DH type II bursts (SE =169.73 kHz)\\
\indent We also examined the bandwidth of DH type II bursts for impulsive and gradual flares associated events. It is directly computed as the difference of starting and ending frequencies (BWD = f$_s$ -- f$_e$). The maximum bandwidth represents that the DH type II bursts covered the entire observing frequency range and it is associated with energetic CMEs (Prakash et al. 2012b). This is a rough estimation of shocks, and radio activities in the IP medium, even it is not continuous. The median bandwidths of DH type II bursts for both sets of events are almost similar (12000 kHz). The difference in the mean bandwidths of DH type II bursts for impulsive and gradual flares events (10285.69 and 10647.46 kHz, respectively) is statistically insignificant.\\
\indent The distributions of duration (a), log peak intensity (b), and slope (c) of DH type II radio bursts for both sets of events are shown in Figure 7. The DH type II bursts associated with shocks are generally detected by the RAD1 and RAD2 at lower frequencies as strong and weak intermittent features. As seen in Figure 7a, the durations of DH type II bursts for all SEP associated events are distributed normally. The mean duration of DH type II bursts associated with impulsive flare events (856.87 min; SE = 110.60 min) is shorter than the gradual flare associated events (1027.23 min; SE = 140.82 min). However, the difference in the mean durations is statistically insignificant. Despite that the median values of one set of events is two times larger than the other set. Many (54\%) of the DH type II bursts? duration of all SEP associated events lies within an hour.  Respectively, 39\% and 59\% of DH type II bursts for impulsive and gradual flare associated events have duration less than 3 hours. From this result, it is noted that gradual flares associated DH type II bursts last longer than the impulsive flare associated events.\\
\indent The distribution of log peak intensity of DH type II bursts is plotted in Figure 7b. It ranges from 2.33 to 5.37 sfu for both sets of events.  As seen from this figure, mean log peak intensity of DH type II bursts is almost equal for both sets of events (~3.64 sfu). We also examined the differences in the integrated intensity of DH type II radio bursts for impulsive and gradual flares associated events. The log integrated intensity ranges from 6.28 sfu-min to 8.72 sfu-min for impulsive and gradual flares associated DH type II bursts. On an average the magnitudes of median log integrated intensity of DH type II bursts are also similar for both sets of events (~7.31 sfu-min). The difference in the mean values is statistically insignificant (P = 74\%). Figure 7c presents the distributions of the slope of DH type II burst. In general, DH radio bursts? slope is directly related to the shock speed. The derived slope of DH type II bursts varies from 0.16 MHz h$^{-1}$ to 1.91 MHz h$^{-1}$. Recently, Winter et al. (2015) showed that the slope of general DH type II bursts for a set of 123 events ranges from 0.2 to 4.6 MHz h$^{-1}$. As seen from this figure, the slope of DH type II bursts for impulsive and gradual flares associated events are similar. The difference in the mean slope of DH type II bursts for impulsive and gradual flares (0.72 MHz h$^{-1}$ and 0.71 MHz h$^{-1}$, respectively) is also statistically insignificant. From these results, we infer that impulsive and gradual flares associated DH type II bursts have similar properties except slight difference in the duration.\\

\begin{figure*}[!htp]
  \includegraphics[width=1.0\textwidth]{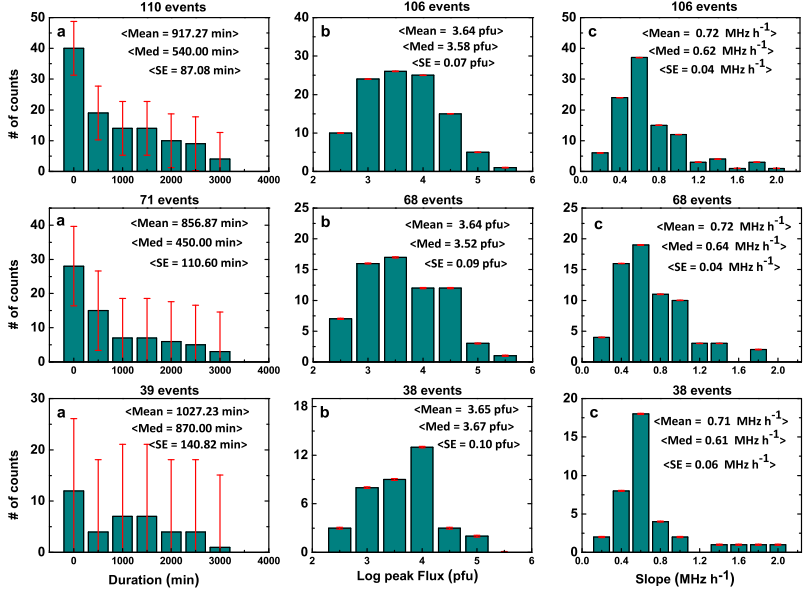}
\caption{Distributions of: (a) duration; (b) log peak intensity; and (c) slope of DH type II radio bursts associated with all SEP events (top), impulsive flares (middle) and gradual flares (bottom) events. The error bars denote the standard statistical error (SE).}
\label{Figure 7}       
\end{figure*}

\subsection{Characteristics of DH type III bursts}

The distributions of duration of DH type III bursts are plotted in Figure 8a and the events are distributed widely from 3 to 62 min. The duration of 14\% of DH type III bursts occurred within 15 min and the peak of this distribution lies at 30 min.  However, the impulsive flare associated DH type III bursts last longer than gradual flare associated events. Interestingly, the mean difference in the mean duration of DH type III bursts for impulsive flare and gradual flare associated events (31.30 min; SE = 1.21 min and 25.08 min; SE = 2.17 min, respectively) is statistically significant (P = 0.1\%). Almost 93\% and 74\% of DH type III durations lie larger than 15 min for impulsive flare and gradual flare associated events, respectively. 
The distributions of log peak intensity of DH type III bursts are presented in Figure 8b. The peak intensity of 69\% DH type III bursts occurred below 1 MHz and for the remaining 31\% events, peak intensity occurred above 1 MHz. The log peak intensity of all SEP associated DH type III bursts varies from 3.87 sfu to 7.57 sfu (median = 6.30 sfu). The median log peak intensity of the DH type III bursts for impulsive flare associated events (6.51 sfu; SE = 0.08 sfu) is relatively larger than gradual flare associated events (5.86 sfu; SE = 0.10 sfu). The difference in the mean values (6.43 sfu and 5.85 sfu,) is statistically significant (P $<< $ 1\%). Recently, \cite{Winter2015} found that the mean log peak intensity of a different set of events (7.10 sfu) is slightly larger than that of all SEP associated DH type III bursts (6.23 sfu).\\
\indent As we already discussed earlier, the importance of DH type III radio bursts accompanying with DH type II bursts in relation to SEP events has been pointed out in the recent studies. For example, some authors discussed about the properties of complex DH type III bursts associated with SEP events in the last two decades \cite{Reiner2000} \cite{Cane2002}. \cite{MacDowall2003} \cite{MacDowall2009} pointed out the occurrence of complex DH type III burst for particle acceleration of SEP events. The results from these studies were thought to imply that SEP originated from the impulsive solar flare acceleration mechanism. But, \cite{Cliver2009} demonstrated that the properties of a set of complex DH type III bursts are not good discriminator for impulsive (from solar flares) and gradual (CME-driven shocks) SEP events. \cite{Gopalswamy2010} also compared three complex DH type III bursts associated with SEP events and found that the occurrence of a DH type III burst is not a sufficient condition for SEP events. Despite these, Winter et al. (2015) found that the result of 92\% of DH type III bursts accompanying with DH type II bursts is a good indicator of SEP events. There are several studies on the duration of complex DH type III bursts observed at 14 MHz and 1 MHz signal (see also., \cite{MacDowall2003} \cite{Gopalswamy2012b}, \cite{Winter2015}. In the present study, we determined the duration of DH type III bursts at 1 MHz signal where DH type III bursts intensity exceeds 6 dB or four times that of the background \cite{MacDowall2003}. We are much aware of the duration of type III burst as responsible for the electron acceleration from the reconnection site. But the long duration of DH type III bursts at 1 MHz signal might also be contributed by the associated CME-driven shocks \cite{Cane1981} \cite{Cane1984}.\\

\begin{figure*}[!ht]
  \includegraphics[width=0.95\textwidth]{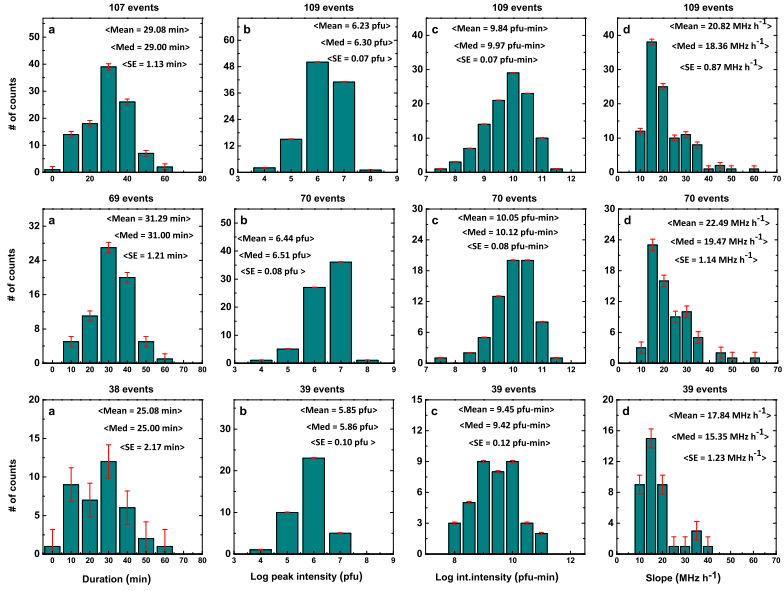}
\caption{Distributions of: (a) duration; (b) log peak intensity; (c) log integrated intensity; (d) slope of DH type III radio bursts associated with all SEP events (top) impulsive flares events (middle) and gradual flare events (bottom ). The error bars denote the standard statistical error (SE).}
\label{Figure 8}       
\end{figure*}

\indent We also derived the integrated intensity of DH type III radio bursts and plotted it in Figure 8c. The distribution for all SEP associated events seems to be symmetric and well fitted with Gaussian behavior. The mean log integrated intensity of DH type III bursts for all SEP associated events is 9.84 sfu-min. The difference in mean log integrated intensity of DH type III bursts for impulsive flare associated DH type III bursts and gradual flare associated events (10.05 sfu and 9.45 sfu, respectively) is statistically highly significant (P$ << $1\%). The distributions of the slope of DH type III bursts are shown in Figure 8d. The average slope of DH type III bursts for impulsive flare associated events (22.49 MHz h$^{-1}$) is larger than that of gradual flares associated events (17.84 MHz h$^{-1}$), and the difference is statistically significant (P = 1\%). From these results, we infer that impulaive flare associated DH type III bursts are relatively longer, stronger and faster than the gradual flares associated DH type III bursts.  We also infered that high energy proton ($>$ 10 MeV) release is mostly (96\%) accompanied with both DH type II and DH type III bursts and this result is well consistant with that of \cite{Winter2015}.

\subsection{Relationships between the properties of SEPs and DH type III bursts}

Many authors have already shown the relationship between the peak intensity of SEPs and CME properties (speeds and widths), and a few authors studied the relationship of SEP intensity with type II burst associated shock speeds. But in this section, we examined the relationships between the key parameters of major SEP events and the propagation of associated DH type III bursts. \\

\begin{figure*}[!ht]
  \includegraphics[width=1.0\textwidth]{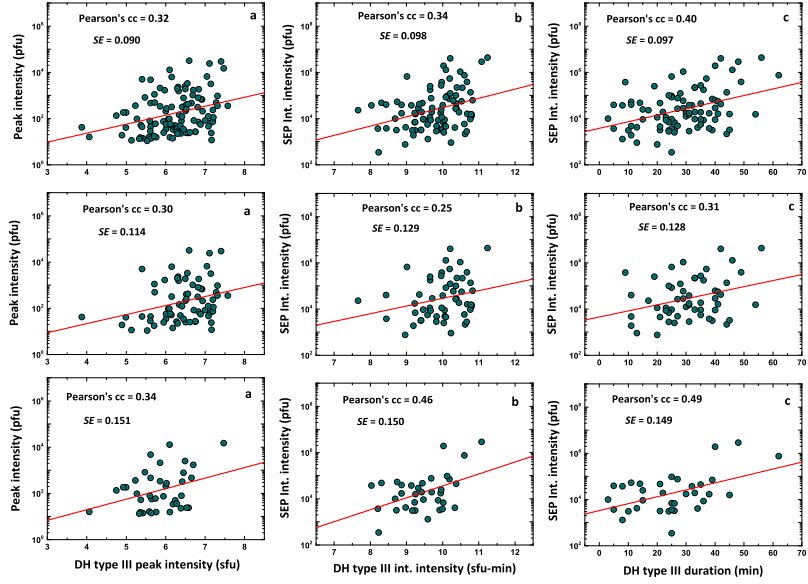}
\caption{Relationship between log peak intensity of SEPs and (a) peak intensity, (b) integrated intensity, and (c) duration of DH type III bursts for all major SEPs (top), impulsive flares events (middle) and gradual flares events (bottom). Linear fitted functions for data points are shown in red line and the corresponding Pearson?s correlation coefficients cc and SEcc are included in each panel.}
\label{Figure 9}       
\end{figure*}

\indent The correlation between the log peak intensity of SEP events and peak intensity of DH type III bursts is shown in Figure 9a. As shown in Figure 9a-top, we found a weak positive correlation between log peak intensity of SEPs and log peak intensity of DH type III bursts (cc = 0.32; SE$_{cc}$ = 0.090). We also plotted the same relationships for impulsive and gradual flares associated events, separately in Fig 9a-middle and bottom rows. Similarly, the relations were examined between the integrated intensity of SEPs and the properties of DH type III bursts (integrated intensity and duration).The correlation for integrated intensity of SEPs and integrated intensity of the DH type III bursts is shown in Figure 9b. We found a better correlation for gradual flare associated events (cc = 0.46; SE$_{cc}$ = 0.150) than the impulsive flare associated events (cc = 0.25; SE$_{cc}$ = 0.129). As seen from Figure 9c-top, we observed a weak positive correlation between the integrated intensity of SEPs and duration of DH type III bursts for all SEP associated events (cc = 0.40; SE$_{cc}$ = 0.097). After splitting these events into impulsive and gradual flares associated events, we found the similar tendency between these properties. The correlation coefficient decreases for the impulsive flares associated events (cc = 0.31, SE$_{cc}$ = 0.128), on contrary the correlation coefficient increases for gradual flare associated events (cc = 0.49, SE$_{cc}$ = 0.149). These results suggest that the DH type III bursts properties (peak intensity, integrated intensity and duration) do also contribute for the peak intensity and integrated intensity of the SEP events. From these results, it is also evident that the enhancement of peak intensity of the SEP events is more closely related with the properties of DH type III radio bursts, but not with the properties of DH type II radio bursts as reported by \cite{Kouloumvakos2015}.   

\section {Summary and conclusions}

In this paper, we have statistically analyzed 115 major SEP events (E$_{p}$ $>$ 10 MeV) and their associated solar flares, CMEs and DH type II bursts listed in CDAW\'s data center during a longer period from November 1997 to December 2014 covering both solar cycles 23 and 24.We have divided them into two sets of events based on the duration of associated solar flares: i) impulsive solar flare associated events (duration $< $60 min) and ii) gradual solar flare associated events (duration $>$ 60) with the aims of finding (i) the important properties of all the associated activities that contribute to the intensity of SEP events and (ii) the distinction between the two sets of events. Almost 14\%, 32\% and 54\% of major SEPs associated events? source locations are distributed respectively in the eastern, disk center and western sides. We have derived the key parameters such as peak intensity, integrated intensity and slope of the DH type II and DH type III bursts from the radio dynamic spectrum. In addition to these, we also estimated the duration and integrated flux of the all major SEP events. The main results are summarized as follows.\\
\indent The mean peak flux and integrated flux for impulsive flare events (X2.9 and 0.32 J m$^{-2}$ respectively) are stronger than the gradual flares (M6.8 and 0.24 J m$^{-2}$, respectively). On an average, the major SEP associated CMEs are highly decelerated (--14.99 m s$^{-2}$) with a mean space speed (1776.66 km s$^{-1}$) three times larger than the non-SEP associated CMEs. The impulsive flare associated CMEs are highly decelerated (--27.66 m s$^{-2}$) with large initial acceleration than the gradual flare associated CMEs (6.26 m s$^{-2}$). This result is consistent with the larger initial speed of impulsive flare-associated CMEs when compared with gradual flare-associated events. The peak speed height for the impulsive flares associated CMEs is slightly lower than the gradual flare associated CMEs. The space speed of CMEs and peak intensity of SEPs for impulsive flare associated events (2275 km s$^{-1}$ and 4568 pfu, respectively) are larger than gradual flare associated events (1881 km s$^{-1}$ and 1198 pfu, respectively). \\
\indent The peak intensity and duration of all 115 major SEP events are 2.18 pfu and 3.04 days, respectively. These results implied that major SEP events seem to be gradual and long duration events. However, the impulsive flare associated SEP events are short-lived and they attained peak intensity relatively quicker (high impulsiveness) than gradual flare associated SEP events. From the correlation studies, we found good correlations for i) peak intensity of SEPs and space speed (cc = 0.52; SE = 0.083) and ii) peak intensity of SEPs and integrated flux of solar flare (cc = 0.44; SE = 0.083). These results imply that both the solar flares and CMEs contribute for the peak intensity of the major SEP events.  After splitting these events into impulsive and gradual flare associated events, the Pearson?s correlation coefficient increased further for impulsive flare-associated events, but it decreased for gradual flare-associated events. This correlation result suggests that there is no clear cut distinction between flare-related and CME-related SEP events for this set of major SEP events. \\
\indent Most of the DH type II bursts properties (peak intensity, integrated intensity and slope) are found almost similar for both the sets of events. But the shock strength of impulsive flare associated DH type II burst events decayed at lower heights than gradual flare associated DH type II burst events. This above result was inferred from the lower ending frequency and longer duration (550.21 kHz and 1027.23 min, respectively) of DH type II bursts for gradual flare associated events. We found that the mean duration, peak intensity, integrated intensity and slope of the DH type III bursts for impulsive flare associated events are longer than gradual flare associated DH type III bursts, and the differences are also found statistically significant. From these results, we infer that impulsive flare associated DH type III bursts are relatively longer, stronger and faster than the gradual flare associated DH type III bursts. It is found that high energy proton release is mostly (96\%) accompanied with both DH type II and DH type III bursts.\\
\indent The positive correlation between the properties of all major SEPs and the properties of DH type III bursts is also good indicator of the enhancement of peak intensity of the SEP events to be more closely related with the properties of DH type III radio bursts than the properties of DH type II radio bursts for this set of events.\\


\begin{acknowledgements}
We thank the referees for their useful constructive comments to improve the quality of this manuscript. We greatly acknowledge the data support provided by various online data centers of NOAA and NASA (CDAW?s team). We would like to thank the Wind/WAVES and RSTN spectrograph teams for providing the type II catalogs. The SOHO/LASCO CME catalog is generated and maintained at the CDAW Data Center by NASA and The Catholic University of America in cooperation with the Naval Research Laboratory. SOHO is a project of international cooperation between ESA and NASA. O. Prakash thanks the Chinese Academy of Sciences for providing General Financial Grant from the China Postdoctoral Science Foundation.
\end{acknowledgements}

%
 \section*{Conflict of interest}
The authors declare that they have no conflict of interest.



\end{document}